\documentclass[a4paper]{article}

\usepackage{INTERSPEECH2020}
\usepackage{tabularx}

\title{MLNET: An Adaptive Multiple Receptive-field Attention Neural Network for Voice Activity Detection}
\name{Zhenpeng Zheng, Jianzong Wang\sthanks{Corresponding author: Jianzong Wang, jzwang@188.com},    \ \ Ning Cheng, Jian Luo, Jing Xiao}
\address{Ping An Technology (Shenzhen) Co., Ltd.}
\email{\{zhengzhenpeng479,\ wangjianzong347,\ chengning221, luojian304,\ xiaojing661\}@pingan.com.cn} 
\begin{document}

\maketitle
 
\begin{abstract}
 Voice activity detection (VAD) makes a distinction between speech and non-speech and its performance is of crucial importance for speech based services. Recently, deep neural network (DNN)-based VADs have achieved better performance than conventional signal processing methods.  The existed DNN-based models always handcrafted a fixed window to make use of the contextual speech information to improve the performance of VAD. However, the fixed window of contextual speech information  can't handle various unpredictable noise environments and highlight the critical speech information to VAD task. In order to solve this problem, this paper proposed an adaptive multiple receptive-field attention neural network, called MLNET, to finish VAD task. The MLNET leveraged multi-branches to extract multiple contextual speech information and investigated an effective attention block to weight the most crucial parts of the context for final classification. Experiments in real-world scenarios demonstrated that the proposed MLNET-based model outperformed other baselines.
\end{abstract}
\noindent\textbf{Index Terms}: Voice Activity Detection, Adaptive Mutiple Receptive-field Attention

\section{Introduction}
Voice Activity Detection (VAD), which aims at removing noise or silence from the original speech  signal and obtaining valid speech signal, is an essential part of speech recognition or other speech-based applications \cite{ax,vf,ctcbased}. Unfortunately, in the real environment, the speech signals may contain numerous background noises and have a low signal-to-noise ratio (SNR), which brings great challenges to an accurate VAD system.

With the continuous development of speech technologies, the research on VAD has become a continuing hot spot \cite{mla,SP1,SP2}.  Early research focused on parametric methods: energy function, zero-crossing rate, statistical signal analysis or other acoustic features \cite{Co,en,Sohn,me}. Later, various machine learning based methods were established to VAD systems: Gaussian Mixture Models (GMM)  \cite{ibm}, Hidden Markov Model(HMM) \cite{hmm} or Support Vector Machines(SVM) \cite{svm}. More specifically, deep learning models were also established: deep neural network(DNN)\cite{mul,Bo2,auc, eva,diff}, deep belief network(DBN) \cite{Zh}, convolutional neural network (CNN) \cite{Ar,te}, recurrent neural network (RNN) \cite{re, acam}. Moreover, speech enhancement (SE) based methods were also introduced to reply the low SNR noisy environments \cite{NC,joint}. 

The above machine learning based models have achieved great progress in VAD task. These models had a common characteristics: in order to get a robust VAD system, speech including contextual information,  just as other speech-based systems, would be forwarded into a neural network. However, when training or testing, it was hard to select the optimal hyper-parameter to determine the amount of contextual information. Specially,  when selecting more information, much noisy information may be included in short speech segments and cause false positive detections. When less contextual information was selected, the VAD system may not make use of effective information to make the right classification. Based on the above analysis, when performing VAD tasks, the VAD systems should be capable of selecting the most appropriate contextual information according to the characteristics of current speech and focusing on the most important speech segments to obtain the optimal detection results. Motivated by the successful application of attention mechanism in image and natural language understanding tasks,  this paper proposed an attention-based model of MLNET to choose the appropriate speech segment for the classifier. The MLNET took advantage of different gated units to extract  different contextual speech information and leveraged attention mechanism of channel selection to choose the most appropriate contextual information to adapt different noisy environments.  In particular, the first attention model for VAD task was the ACAM model and it focused on the effect of a certain frame in a fixed window, which is different from our window size attention \cite{acam}. From our point of view, the most important frame for the detection result is the current frame and the surrounding frames are also important for final result but they are the auxiliary information.

With respect to the state of the art, the main contributions of this paper were as following:  
\begin{itemize}
\item A architecture of MLNET, which could adapt different segments and select the optimal contextual information for the final classifier, was proposed to VAD.
\item A useful mechanism of gated units was leveraged to extract different contextual speech information.
\item An attention strategy for effective and appropriate contextual information selection was investigated. 
\item The proposed method was benchmarked against several state-of-art methods and the functions of each part in MLNET were also compared.
\end{itemize}

\section{Methodology}
\subsection{Model Structure}
The MLNET-based VAD achieved a frame-based speech or non-speech classifier. Suppose the training corpus can be marked as $\chi={\{(x_t,y_t)\}}_{t=1}^T$ , where $x_t$ is an acoustic feature vector of $t$-th frame audio signal and $y_t$ denotes the label of $x_t$. If $x_t$ is a speech frame, then $y_t=1$; otherwise, $y_t=0$. Because of contextual information's importance for speech applications, $x_t$ is usually expanded to $I_t=[x_{t-r},...,x_{t-1},x_t,x_{t+1},...,x_{t+r}]$ when training or testing. Here, the objective of VAD is to learn a function $f$ as (1).
\begin{equation} 
\hat{y_t} = f([x_{t-r},...,x_{t-1},x_t,x_{t+1},...,x_{t+r}]) 
\end{equation}
where $\hat{y_t}$ denotes the predicting result of $x_t$ and $r$ denoted the window size of contextal speech information.

In experiment, the value $r$ was a hyper-parameter and the fixed value of $r$ was hard to adapt various speech environments. For example, when $x_t$ wa a speech frame and the speech segment duration around $x_t$ was short, the large $r$ would contain much non-speech information and cause false positive results. In turn, when $r$ was short, the contextual information may not be fully utilized, which can result in false negative or false positive results. To address the above problems, this paper proposed the MLNET model, which leveraged the multiple gated affined units and attention mechanism to choose the optimal receptive-field speech adaptively to make the classification. The MLNET's architecture is shown in Figure 1.
 
\begin{figure}[htbp] 
\centerline{\includegraphics[width=0.23\textwidth,height=0.22\textheight]{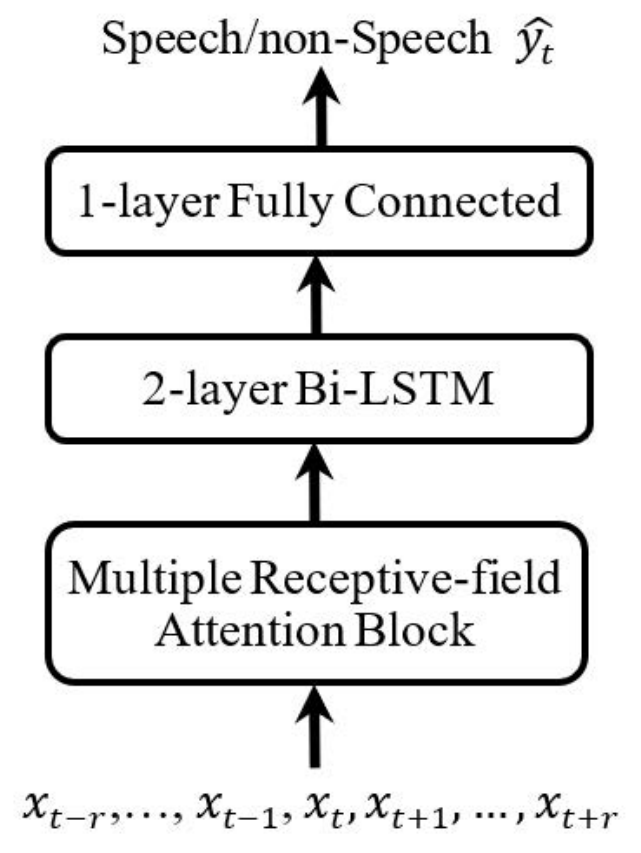}}
\caption{The MLNET's Architecture. The $x_{t-i}(i\in [-r, +r])$ represented 40-log Mel Features of a frame and $(2r + 1)$  frames' features would be inputted into the MLNET.} 
\end{figure}

\subsection{Adaptive Multiple Receptive-field Attention Block}


The detail architecture of multiple receptive-field attention block was showed in Figure 2. In order to realize adaptive feature selection, multiple branches were leveraged to extract different receptive-filed speech features and each branch represents a specific receptive field or a specific contextual information. Specifically, 
for the branch of $r_i$, the contextual speech information can be denoted as $I_t^{'}=[x_{t-r_i},...,x_{t-1}, x_t, x_{t+1},...,x_{t+r_i}]$  and $(2\times r_i + 1)$ frame features were included.  
For the calculation of subsequent attention modules, the feature matrix of $I_t^{'}$ with different size need to be converted feature matrix with the same size. This paper made use of the gated affined unit to finish this task.  The gated affined unit as this non-linearity has proved to work better for modeling audio than the other affine function or Relu functions \cite{gated}.  
 The gated affined unit's definition was defined as (3):
\begin{equation}
q_i=tanh(W_{f,r_i}*I_i + b_{f, r_i})\odot\sigma(W_{g,r_i}*I_i + b_{g, r_i})
\end{equation}

where $W_{f,r_i}$, $b_{f, r_i}$ denoted affine parameters of tanh and $W_{g,r_i}$, $b_{g, r_i}$ denoted affine parameters of sigmoid.  It was obvious that the size of $W_{f,r_i}$ and $W_{g,r_i}$ would change with $r_i$ while the size of $q_i$ was constant regardless of $r_i$. The $*$ denoted a convolution operator while the $\odot$ denoted an element-wise multiplication operator.

\begin{figure}[htbp] 
\centerline{\includegraphics[width=0.45\textwidth,height=0.25\textheight]{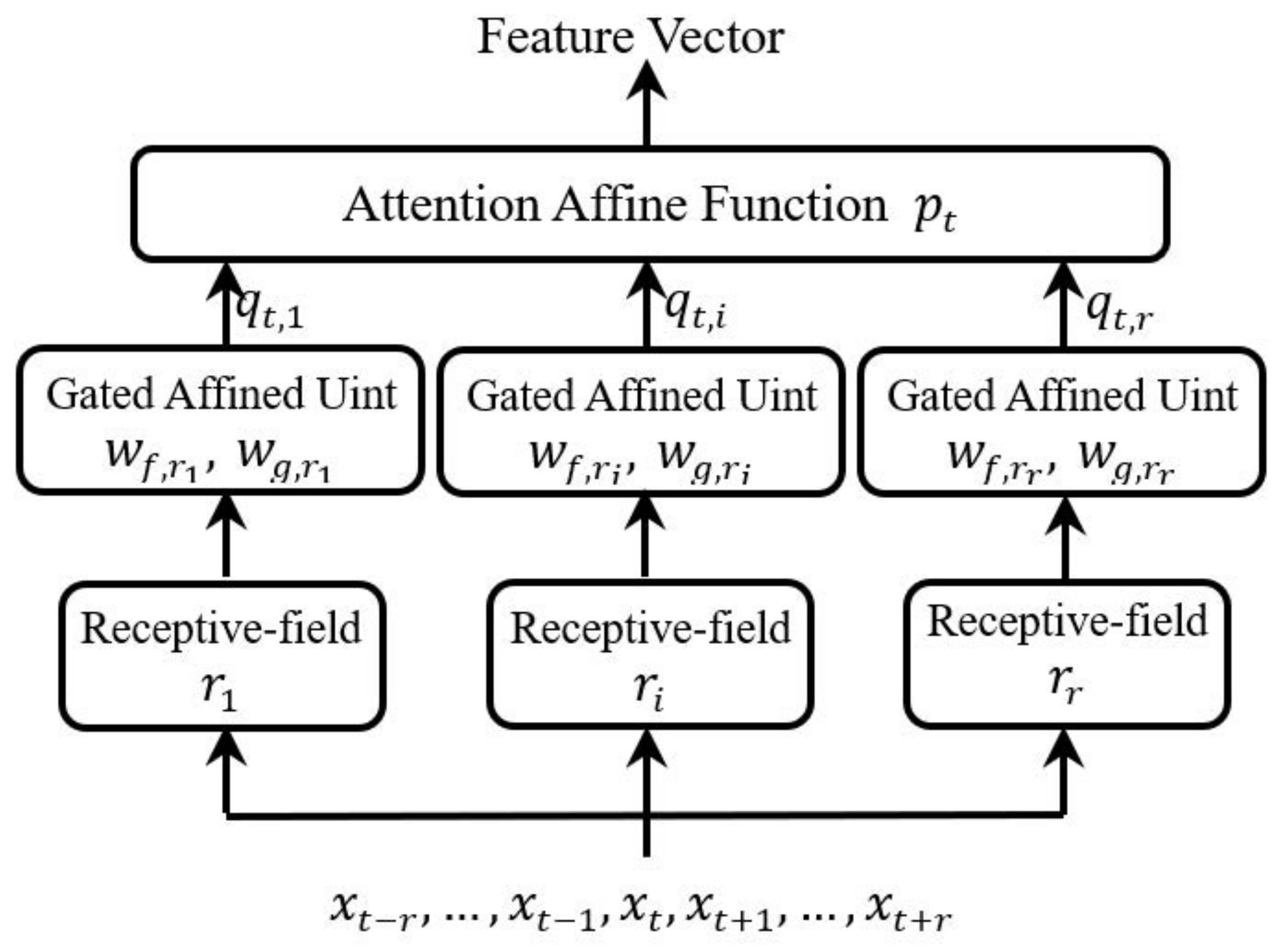}}
\caption{Adaptive Multiple Receptive-field Attention Block. As can be seen, the inputted feature with different receptive-field $r_i$ would be calculated at different branches and the attentional feature vector of time $t$ would be calculated based on each branch.}
\end{figure}

After gated affined unit operation, different two-dimensional extracted matrixes, which were denoted by $q_{t,1},...,q_{t,i},...,q_{t,r}$,  were obtained and each feature matrix of $q_{t,i}$ had the same size.  Analogy with image, each feature matrix of $q_{t,i}$ can be regarded as a channel feature map and we produced a channel attention map to decide which feature matrix should be focused on. Based on previous channel attention work in images \cite{senet,cbam}, the attention module's structure was showed in Figure 3. Firstly, both average-pooling and max-pooling operations were used to aggregate information of a feature matrix and two feature descriptors were generated to represent different feature matrixes. Then, both descriptors were forward to a shared 2-layer fully connected DNN to produce two different attention ratio vector $p_{t,max}$ and $p_{t,avg}$. Finally, by adding an norm operation, $p_{t,max}$ and $p_{t,avg}$ were used  to generate final attention ratio vector $a_t$ of (4).  
\begin{equation} 
\begin{aligned}
a_{t} &= \sigma{(FC(avgpool(q_t)) + FC(maxpool(q_t)))} \\
&=\sigma{(W_1(W_0(avgpool(q_t))) + W_1(W_0(maxpool(q_t))))}
\end{aligned}
\end{equation}
where $FC$ denoted the shared 2-layer fully connected DNN and $W_1$ and $W_0$ were the weights of DNN. Note that the weight $W_0$ of first layer followed a $leaky\_relu$ activation function. After obtaining $a_t$, the attentioned or scaled feature matrix would be calculated through (4)-(7).
\begin{equation}
\begin{aligned}
a_t &= [a_{t,1},...,a_{t,i},...,a_{t,r}]
\end{aligned}
\end{equation}
\begin{equation} 
p_{t,i} = \sigma{(a_{t,i})} / (\sum_{i=1}^{r}{\sigma{(a_{t,i})}})
\end{equation}
\begin{equation} 
p_{t} = [p_{t,1},p_{t,2},...,p_{t,r}]
\end{equation}
\begin{equation} 
Q_{t} = \sum_{i=1}^{r}{p_{t,i}*q_{t,i}}
\end{equation}

Where $a_t$ denoted the attention vector and it can be calucated by the attention module. 
$\sigma$ was the sigmoid function and the $p_t$ represented the normalized value of  $a_t$. Finally, the $Q_t$ was calculated by summing original $q_t$ with attentional coefficient.

\begin{figure}[htbp] 
\centerline{\includegraphics[width=0.24\textwidth,height=0.26\textheight]{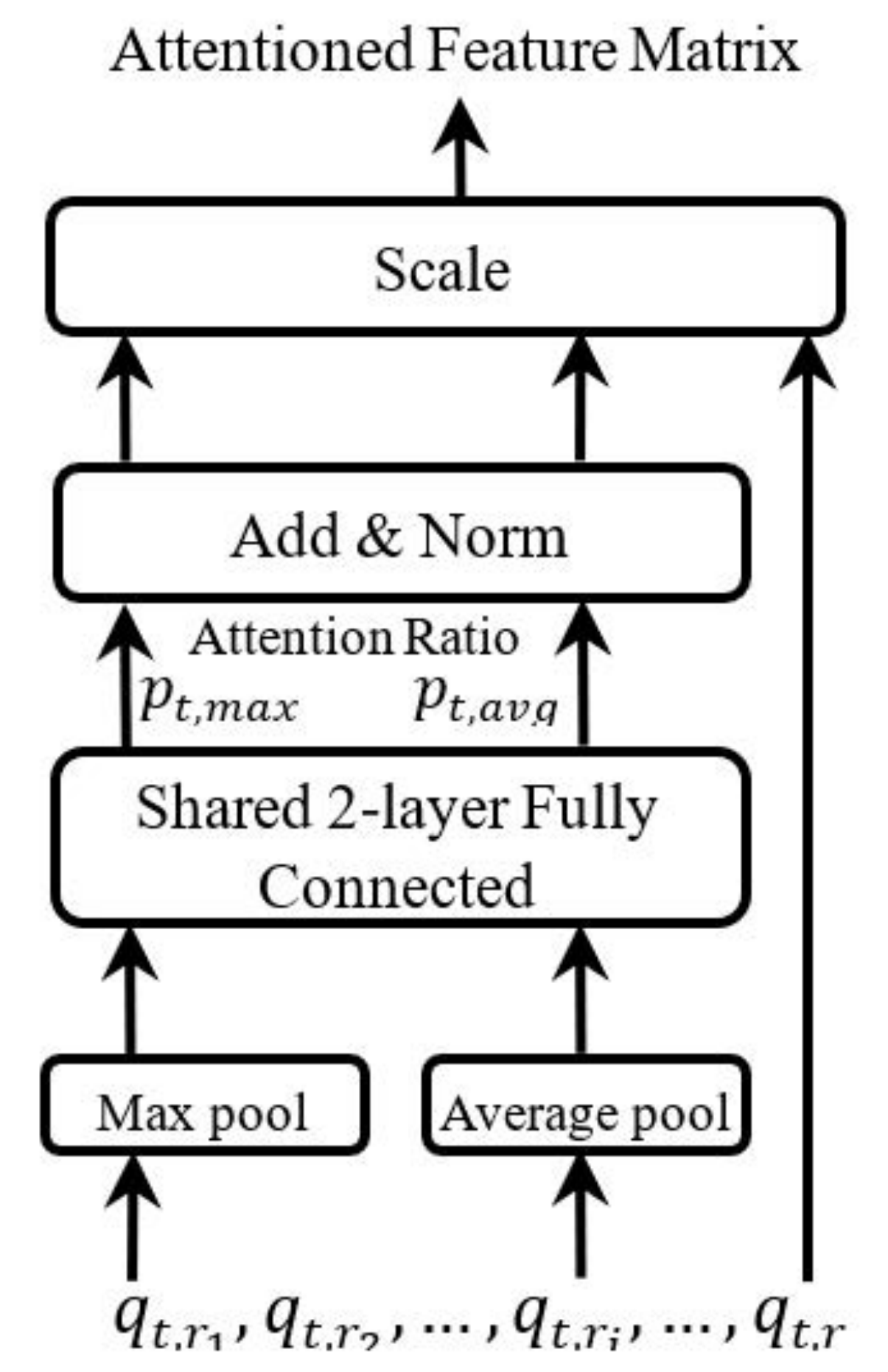}}
\caption{Attention Module Structure. The gated affined unit $q_i$ of receptive-field $r_i$ would produce two feature vector through maxpooling and averagepooling and the two vectors were inputted 2-layer FC to produce the attentional weights.  The final attentional feature matrix can be obtained through (5)-(8)}
\end{figure}

\subsection{Augmented Classifier}
Through the adaptive multiple receptive-field attention block, the scaled feature matrix $Q_t$ was obtained and $Q_t$ would be reshaped into a feature vector $m_t$. The Bi-LSTM is skilled in learning contextual speech information and the DNN is an excellent classifier.  Nextly, the feature sequence $M_t = [m_1,m_2,...,m_T]$ would be fed into a two-layer Bi-LSTM and 1-layer Fully Connected DNN to make the final classification. 

The training of MLNET-based VAD could be regarded as a common supervised optimized problems with traditional cross-entropy loss function. 
\begin{equation}
\mathcal{L}_{crossentropy} = \sum_{t=0}^{t=T}{\{y_{t}log(\hat{y}_t) + (1-y_{t})log(1-\hat{y}_t)\}}
\end{equation}

Because of the characteristic of the proposed model, this paper further investigated an additional attention loss function to adapt the attention mechanism in the training phase. The attention loss function was designed to emphasize the most important receptive field of  $r_k$, and the definitions were showed in (9)-(11).
\begin{equation}
k = \underset{i}{argmax}(p_{t,i}) \\
\end{equation}
\begin{equation}
\mathcal{L}_{attention} =  \sum_{t=0}^{t=T}\sum_{i=0}^{i=r}(y_{t,i}log(p_{t,i}))
\end{equation}
\begin{equation}
\mathcal{L} = \mathcal{L}_{crossentropy} + \mathcal{L}_{attention}
\end{equation}
Where $ \mathcal{L}_{attention}$ could be denoted as a softmax problem and the most appropriate receptive-field $p_{t,k}$ was the target and the corresponding $y_{t,k}$ is $1$ and other $y_{t,i} = 0({i\not=k})$.

\section{Experiments}
\subsection{Datasets and Evaluations}
In this paper, the English corpus of Aurora4 \cite{aurora4} and the Chinese corpuses of Thchs30 \cite{ths30} were applied to train and test the proposed model. In our experiments, firstly, because of imbalance of speech and non-speech,  2-second-long silence segments were added into forward and backward of each utterance. In training,  the clean speech corpus were corrupted by public 100 noise types of HuNonspeech\footnote{http://web.cse.ohio-state.edu/pnl/corpus/HuNonspeech/}. Each utterance was randomly corrupted at a level of -5dB-20dB SNR and all  have an average 7.5dB SNR. The NOISEX-92\footnote{http://spib.rice.edu/spib/select\_noise.html} noise dataset was used to construct testing dataset and 4 types unseen noises of babble, factory, destory-engine were selected to corrupt the clean speech. Also, the SNR was setted between -5dB and 20dB and the average is 7.5dB.  For Aurora4, 95\% of training data were used for training and 5\% were used as dev data. The testing corpus of Aurora4 were corrupted by the NOISEX-92 and leveraged as the testing data.  For thchs30 data, the dataset was constructed with the same process, but the different was that dev dataset leveraged the original corresponding data.

 For comparison, the metrics of F1-score and DCF were selected as a performance measurement. F1-score took into account both accuracy and recall metrics, which was a common evaluation index of binary classification problems. DCF\footnote{http://fearlesssteps.exploreapollo.org/}  reflected the wrong performance of the model and DCF was defined as followed:
\begin{equation}
F1-score(\theta) = \frac{2TP}{2TP + FP + FN}
\end{equation}
\begin{equation}
DCF(\theta) = 0.75 \times P_{FN}(\theta) + 0.25 \times P_{FP}(\theta)
\end{equation}
where $\theta$ denoted a given system decision-threshold setting. TP represented true positive examples' num while FP and FN represented the num of false positive and false negative examples.
$P_{FN}$ was the probability of FP while $P_{FN}$ was the probability of FN. It should be noted that the larger the F1-score was, the better performance while the smaller the DCF was, the better performance. In testing, we calculated the two metrics of each recording respectively and averaged the metrics of all recordings as the final score.


\begin{table}
\centering 
\caption{Model Configuration}
\begin{tabularx}{7cm{\centering}}{lXl<{\centering}}  
\hline
\hline
Name  &Unit \\ 
\hline 
\#Attention module \\  
shared 2-layer FC & 64$\times$64 \\   
\hline
\#Main network \\    
Affine Matrix &40$\times (2*[1,3,5,7,9]+1)$ \\  
              &40$\times (2*[1,3,5,7,9]+1)$ \\  
Attention module  &1 &  \\ 
2-layer Bi-LSTM  &(64 + 64) $\times$ (64 + 64) \\
1-layer FC  &64  \\    
\hline 
\hline
\end{tabularx}  
\end{table}

\subsection{MLNET Setup}
The acoustic feature extracted for MLNET was 40-dimensional log-mel filterbank while the frame size was 25 ms with a shift of 10 ms. The window of attention block were setted as 19 frames, which corresponded to 190ms contextual information, while the gated affined unit's receptive-field were setted 1, 3, 5, 7 ,9. Other parameters of MLNET were shown in Table1. For training our proposed model, the matrix weight parameters of  MLNET were all initialized with random uniform initialization  while the bias parameters were initialized with a constant of 0.1. In our experiments, we trained the network for 150 epoches with the Adam algorithm when the loss function got little change. The batchsize was 32 and the learning rate was set to 0.001. When training, the gradient cropping strategy was also applied and the gradient of each parameter at each iteration was limited between -1 and 1. 

To demonstrate the effectiveness of our proposed model, three VAD approaches were used for comparison. The first was google' WebRTC VAD systems \cite{google}. Additionally, Vafeiadis proposed 2-D CRNN model and has made great success in recent speech activity detection \cite{crnn}. Nextly, ACAM  approach, which was the state of art of attention-based methods, was also included \cite{acam}.   In our experiments, we made use of the same parameters, but all trick strategies, such as batch normalization and regularization, were not leveraged. To alleviate the effect of the input features, it should be noted that 40-log mel acoustic features were leveraged to establish the CRNN and ACAM baseline models, which were different from the original approaches. 

\begin{table} 
\centering 
\caption{Result Comparision of Aurora4}  
\begin{tabularx}{6cm{\centering}}{lXl<{\centering}}  
\hline
\hline
Name  &Dev &Eval\\ 
\hline
\#\textbf{F1-score} \\
Google VAD (mode 0) & 72.33 &76.32 \\   
CRNN &89.14 &87.23 \\ 
ACAM &90.56 &89.03   \\    
\textbf{MLNET} &\textbf{91.38} &\textbf{89.27}\\
\hline
\#\textbf{DCF} \\
Google VAD (mode 0) &22.06 &18.34  \\     
CRNN &9.23 &10.34  \\     
ACAM &8.95 &9.01 \\    
\textbf{MLNET} &\textbf{8.77} &\textbf{9.23}\\
\hline 
\hline 
\end{tabularx}  
\end{table}

\begin{table} 
\centering 
\caption{Result Comparision of Thchs30}  
\begin{tabularx}{6cm{\centering}}{lXl<{\centering}}  
\hline
\hline
Name  &Dev &Eval\\ 
\hline
\#\textbf{F1-score} \\
Google VAD (mode 0) &74.71 &74.60  \\     
CRNN &91.35 &89.90  \\     
ACAM  &92.53 &91.27 \\    
\textbf{MLNET} &\textbf{93.25} &\textbf{92.58}\\
\hline
\#\textbf{DCF} \\
Google VAD (mode 0) &17.88 &18.18  \\     
CRNN &8.21 &9.72  \\     
ACAM  &7.67 &8.51 \\    
\textbf{MLNET} &\textbf{6.89} &\textbf{8.12}\\
\hline 
\hline 
\end{tabularx}  
\end{table}

\subsection{Results and Discussions}
The results were summarized in Table 2 and Table 3. We can observed that MLNET has the best performance and outformed other three baselines models. Especially, three-training-based methods achieved 10\% higher than google's VAD in F1-score and 8\% lower in DCF. The ACAM and MLNET outformed the traditional deep learning methods of CRNN, which proved attention-based models were helpful for improving the accuracy of detection. Comparing with ACAM,  MLNET selected the windows' attentions instead of ACAM's frames-based attention and experiments showed that the window based attention models achieved higher performance than the frames-based models, which also conformed to our prior knowledge that current frame information was the most important and the joined frames were auxiliary information when predicting the current frame.  


In order to illustrate each part's functionality of our proposed model, the comparison of each modules were further investigated. The base was the Bi-LSTM, which just leveraged the contextual speech information and the feature vectors of  contextual speech were aggregated to a longer vector before feeding into the network. The second was the gated unit model that the gated unit operation replaced the aggregated mechanism. The third and the fourth were leveraged to certificate the multiple window's  functions while the attention's function was also compared. The Aurora4 dataset were leveraged to make this evaluation and the results were shown in Table 4.  As noted in this table, the gated affined unit based models have better performance than direct aggregation of Bi-LSTM and achieved about 2\% increase in F1-score and 1.5\% decrease in DCF.  Adding the multiple receptive-field of non-attention, the VAD's performance was also improved and achieved   about 2.5\% increase in F1-score and 2.13 \% decrease in DCF. Lastly, the adaptive attention mechanism was also helpful for MLNET's performance.  In contrast, the adaptive attention only achieved a small accuracy improvement than non-attention module and the reason may be the mechanism of receptive-field selection has existed in multiple receptive-field attention block. In particular, we observed that the adaptive multiple receptive-field attention block was also helpful for speeding up models' convergence in our experiments. To sum up, the proposed method can better deal with the VAD problems.

\begin{table} 
\centering 
\caption{Module Comparision}  
\begin{tabularx}{6cm}{lXl}  
\hline
\hline
Model  &Dev &Eval\\ 
\hline
\#\textbf{F1-score} \\
Bi-LSTM &85.89 & 84.17 \\     
+Gated Unit &87.63 &86.24 \\     
+Non-Attention &90.85 &88.73  \\    
\textbf{+Attention} &\textbf{91.38} &\textbf{89.27}\\
\hline
\#\textbf{DCF} \\
Bi-LSTM &12.24 &13.54  \\     
+Gated Unit &11.16 &12.01 \\     
+Non-Attention &9.11 &9.88  \\    
\textbf{+Attention} &\textbf{8.77} &\textbf{9.23}\\
\hline 
\hline 
\end{tabularx}  
\end{table}

\section{Conclusion}
The existed DNN-based VAD models only leveraged fixed receptive-field contextual speech information and were unable to handle with speech segments of different lengths adaptively. To overcome this defect, this paper proposed an architecture of MLNET for VAD task. MLNET made use of different gated affined unit to extract different contextual speech information and leveraged the adaptive attention block to select the most focused speech segments. Comparing with the existed models, the experiments demonstrated that MLNET outformed other baseline models and proved that the proposed architecture was helpful to deal with VAD problems.

\section{ACKNOWLEDGEMENTS} 
This paper is supported by National Key Research and Development Program of China under grant No. 2018YFB1003500, No. 2018YFB0204400 and No. 2017YFB1401202. Corresponding author is Jianzong Wang from Ping An Technology (Shenzhen) Co., Ltd.

\bibliographystyle{IEEEtran}
\bibliography{mlnetbibfile}

\end{document}